\documentclass[3p,times]{elsarticle}
\usepackage{ifpdf}
\usepackage[usenames]{color}
\usepackage{graphicx}
\usepackage{amssymb}
\usepackage{amsmath}

\newcommand{\prl}{Phys. Rev. Lett.\ }
\newcommand{\pra}{Phys. Rev. A\ }

\journal{Physics Letters A}

\begin{document}

\begin{frontmatter}

\title{Dipolar Bose-Einstein condensate soliton on a two-dimensional optical lattice}

\author[ift]{S. K. Adhikari}
\ead{adhikari@ift.unesp.br}
\author[ift,bdu]{P. Muruganandam\corref{muru}}
\ead{anand@cnld.bdu.ac.in}
\cortext[muru]{Correspondig author; Telephone: +91 431 2407057; Fax: +91 431 2407093}

\address[ift]{Instituto de F\'{\i}sica Te\'orica, UNESP - Universidade Estadual Paulista, 01.140-070 S\~ao~Paulo, S\~ao Paulo, Brazil}
\address[bdu]{School of Physics, Bharathidasan University, Palkalaiperur Campus, Tiruchirappalli  620024, Tamilnadu, India}

\begin{abstract} 
Using a \emph{three-dimensional} mean-field model we study one-dimensional dipolar Bose-Einstein condensate (BEC) solitons  on  a weak two-dimensional (2D) square and triangular optical lattice (OL) potentials 
placed perpendicular to the polarization direction.  The stabilization against collapse and expansion  of these solitons for a fixed dipolar interaction and a fixed number of atoms is possible  for short-range atomic interaction lying   between two  critical limits. The solitons collapse below the lower limit and escapes to infinity above the upper limit.  One can also stabilize identical tiny BEC solitons arranged on the 2D square OL sites forming a stable 2D array of interacting droplets when the OL sites are filled with a filling factor of 1/2 or less. Such an array is unstable when the filling factor is made more than 1/2 by occupying two adjacent sites of OL.  These stable 2D arrays of dipolar superfluid BEC solitons are quite similar to  the recently studied dipolar Mott insulator  states on 2D lattice in the Bose-Hubbard model by Capogrosso-Sansone \emph{et al.} [B. Capogrosso-Sansone, C. Trefzger, M. Lewenstein, P. Zoller, G. Pupillo, Phys. Rev. Lett. 104 (2010) 125301].
 \end{abstract}

\end{frontmatter}

\section{Introduction}
After the experimental observation of Bose-Einstein condensate (BEC) \cite{rmp},  
there has been great interest in the problem of stabilization 
of BEC on a periodic optical lattice (OL) potential. The OL potential 
is generated in a laboratory by a standing wave   polarized
laser beam \cite{sw}. The resulting periodic potential simulates the potential 
seen by an electron in a solid \cite{band}. 
As one of the interests in studying BEC droplets  of small number of atoms on OL
is to generate a stable array of BEC droplets by occupying each OL site with one 
tiny droplet so that an array of matter wave is formed as in condensed-matter physics  \cite{lewen}.
Unlike in condensed-matter physics, these BEC droplets are completely 
pure. Such a  pure array of matter-wave simulating a quantum solid 
can be created and studied in a laboratory
to model many quantum-mechanical condensed-matter phenomena.

The above study to model condensed-matter phenomena 
with matter wave has obtained new impetus after the observation of 
dipolar 
BEC of $^{52}$Cr \cite{pfau,rpp} and $^{164}$Dy \cite{dy1} with a large 
dipolar interaction. 
The dipolar interaction is anisotropic and 
of long range in contrast to the short-range isotropic atomic interaction. 
Because of the anisotropic long-range interaction, the conditions of stability 
of a dipolar BEC soliton follow a distinct trend from that of a nondipolar 
BEC soliton \cite{stab}. Also, dipolar atoms have permanent 
magnetic moment  \cite{pfau} and if 
polarized by an external magnetic field, an array of tiny BEC droplets 
with magnetic moment 
can simulate the problem of generation of magnetism in solids from 
individual atomic or molecular magnetic dipoles.

Here,   using   
a three-dimensional (3D) mean-field model,
we study
one-dimensional (1D) 
dipolar
BEC  solitons, free to move along the polarization direction $z$, 
on a two-dimensional  (2D) square and triangular 
periodic OL in the $x$-$y$ plane.   
In both cases, for a fixed dipolar interaction and fixed number of atoms, the solitons 
are stable for atomic short-range interaction (scattering length) between two critical 
limiting values. Below the lower limit the dipolar and short-range interactions 
lead to too much attraction 
and the soliton collapses and above the upper limit the net attraction is too weak 
and the soliton escapes to infinity.
The spreading of the BEC soliton in the 2D OL in the
$x$-$y$
plane is stopped by the attractive dipolar 
interaction along polarization direction $z$ $-$ the system
lowers
its energy by being long in the $z$ direction and thin in the $x$-$y$ plane.

First, we consider a  dipolar BEC  soliton    
on a 2D  square or triangular OL
using the numerical and Lagrangian variational analysis  
of the mean-field Gross-Pitaevskii  (GP) equation. 
The variational results are  found to be 
in good agreement with numerical 
results. 
To demonstrate the stability,
we  perform a linear stability analysis \cite{stability} 
  using the variational solution
and calculate the normal-mode frequencies  \cite{luca}.

In an attempt to generate an array of dipolar BEC solitons 
on a 2D square OL, each with a small number of atoms,  
we find that interesting stable periodic structure can be formed for filling factors 
of 1/2, 1/3, and 1/4 termed checkerboard, stripe, and star configurations, respectively.   
Similar stable structures of ultra-cold dipolar atoms 
were first obtained as Mott insulator states
in a study of dipolar atoms on 2D OL by solving the 
corresponding { \it field-theoretic} 
2D Hubbard model numerically by Monte Carlo technique \cite{barbara}. 
This suggests that such stable structures are a consequence of 
the typical repulsive dipolar interaction in the $x$-$y$ plane.

There have been studies of 2D dipolar BEC solitons, 
free to move in the $x$-$z$ or $x$-$y$ plane    with  harmonic traps 
along $y$ \cite{Tikhonenkov2008} or $z$ \cite{Pedri2005} axis, respectively,
and of 1D dipolar BEC soliton  under transverse harmonic trap \cite{adhisol}.
The present 1D dipolar BEC soliton   
confined by only a weak 2D OL in the $x$-$y$ plane
 is distinct. The BEC soliton of the previous studies 
\cite{Tikhonenkov2008,Pedri2005,adhisol} will essentially have a Gaussian 
density distribution along the infinite trap direction, whereas the 
present BEC soliton will have an exponential density distribution due to 
weak finite OL traps in these directions.  Similar nondipolar  BEC solitons in a lower dimensional OL have also 
been studied \cite{salerno}.

\section{Analytical formulation}
We consider  a
1D dipolar BEC soliton  of $N$ atoms, each of mass $m$, using the  GP
equation: \cite{pfau}
\begin{eqnarray}  \label{gp3d}
i \frac{\partial \phi({\bf r},t)}{\partial t}
=  \biggr[ -\frac{\nabla^2}{2} + V_{\text{OL}}^{2D}
+\bar \mu(a,N)
+
\bar\mu_{dd}(a_{dd},N)\biggr] \phi({\bf r},t),
\end{eqnarray}
with the bulk chemical potential 
$
\bar \mu(a,N) = 4\pi a n, \quad n=N|\phi|^2,$
with $a$ the atomic scattering length,  $n$ the  density, and  ${\bf r}\equiv
\{x,y,z \} \equiv \{\rho,z\}$.
The dipolar bulk chemical
potential is
$\bar\mu_{dd}(a_{dd},N)=N\int U_{dd}({\bf r-r})|\phi({\bf r'},t)|^2d{\bf r'}$,
 $U_{dd}({\bf R}) =  3a_{dd}
(1-3\cos^2\theta)/R^3$ the dipolar interaction potential,
 ${\bf R=r-r'},$
 normalization $\int \phi({\bf r})^2 d {\bf r}$ = 1,
   $\theta$
the angle between $\bf R$ and polarization direction  $z$,
$V_{\text{OL}}^{2D}=  -V_0 \cos(2x)-V_0\cos(2y)$ for the square OL, and 
=$ -V_0 \cos(2x)
-V_0\cos(x+\sqrt{3}y)-V_0 \cos(x-
\sqrt{3}y)$ for  the  triangular
OL, with  $V_0$  the strength of the OL.
This 2D triangular OL consists of three OL in the $x$-$y$ plane at mutual angles
of $\pi/3$.  
The length $a_{dd}
=\mu_0\bar \mu^2 m /(12\pi \hbar^2)$
is
the strength of
dipolar interaction,
 $\bar \mu$ the magnetic dipole moment of an   atom, and $\mu_0$
the permeability of free space.
  In  Eq. (\ref{gp3d}),
length is measured in
units of  $l_0 \equiv
\lambda/2\pi$, taken here as 1 $\mu$m,  and time $t$ in units of $t_0 = ml_0^2/\hbar$, where
$\lambda$ is the OL wave length.
Energy $E$ and $V_0$ are expressed in units of $2E_R=h^2/(m\lambda^2)$, 
where $E_R$ is the recoil energy of one atom of mass
$m$ absorbing one lattice photon of wave length $\lambda$.

First we consider a 1D BEC  soliton  on the OL  
$ V_{\text{OL}}^{2D}$. The  
Lagrangian density of Eq.
(\ref{gp3d})    is \cite{jb}
\begin{align}
{\cal L}=& \,\frac{i}{2}\left( \phi \phi^{\star}_t
- \phi^{\star}\phi_t \right) +\frac{1}{2}\vert\nabla\phi\vert^2
+   2\pi aN\vert\phi\vert^4+V^{2D}_{\text{OL}}|\phi|^2
+ \frac{1}{2}N\vert
\phi\vert^2\int U_{dd}({\mathbf r}-
{\mathbf r'})\vert\phi({\mathbf r'})\vert^2 d{\mathbf r}'
.\label{eqn:vari}
\end{align}
For a variational study we
use the Gaussian ansatz  \cite{jb}: 
$ \phi({\bf
r},t)=
\exp(-
{\rho^2}/{2w_\rho^2}- {z^2}/{2w_z^2}$ $ +i\gamma\rho^2
+i\beta z^2 )  /({w_\rho \sqrt w_z}\pi^{3/4})$
where $w_\rho$ and $w_z$ are time-dependent widths and 
$\gamma$ and $\beta$ are time-dependent chirps. 
Because we consider identical strengths of OL along $x$ and $y$ 
directions, an axially-symmetric Gaussian profile for the density 
is a good approximation to the actual density.   
The
effective Lagrangian $L$ (per particle) is
\begin{align}\label{lag}
L \equiv  &\,  \int {\cal L}\,d{\mathbf r} 
 =  \left(w_\rho^2\dot{\gamma} +\frac{1}{2}
w_z^2\dot{\beta}+2w_\rho^2\gamma^2+w_z^2\beta^2 \right) 
+  E_{\mathrm{kin}}+ E_{\mathrm{trap}}+  E_{\mathrm{int}}, 
\end{align} 
with kinetic, trap, and interaction energies given, respectively,  by  
$E_{\mathrm{kin}}=  -[{1}/{(2w_\rho^2)} + {1}/{(4w_z^2)}]$, 
$E_{\mathrm{trap}}= -2V_0 \exp(-w_\rho^2)$, for the square OL 
and $=-3V_0 \exp(-w_\rho^2)$, for the triangular OL, 
${ E}_{\mathrm{int}}= N[a-a_{dd}f(\kappa)]/(\sqrt{2 \pi}w_\rho^2w_z)$, 
where 
$  
f(\kappa)=[1+
2\kappa^2-3\kappa^2$ $d(\kappa)] /
(1-\kappa^2)$, $d(\kappa)
=\mbox{atanh}\sqrt{1-\kappa^2}/\sqrt{1-\kappa^2}$, $\kappa=w_\rho/w_z.$
The Euler-Lagrange equations for parameters 
$ w_\rho, w_z, \gamma, \beta$
can be used to obtain the following equations of
the widths for the dynamics of the dipolar BEC  soliton
\begin{align} \label{f3} 
\ddot{w}_{\rho} & =  
\frac{{{1}}}{w_\rho^3} +\frac{
1}{\sqrt{2\pi}} \frac{N}{w_\rho^3w_{z}}
\left[2{a} - a_{dd}{g(\kappa) }\right]  +2E_{\mathrm{trap}}w_\rho
  ,
\\  \ddot{w}_{z} & = 
\frac{1}{w_z^3}+ \frac{ 1}{\sqrt{2\pi}}
\frac{2N}{w_\rho^2 w_z^2} \left[{a}-
a_{dd}h(\kappa)\right]  , \label{f4} \end{align}
with $g(\kappa)=[2-7\kappa^2-4\kappa^4+9\kappa^4
d(\kappa)]/(1-\kappa^2)^2, h(\kappa)= [1+10\kappa^2-2\kappa^4-9\kappa^2
d(\kappa)]/(1-\kappa^2)^2.$
The widths of a stationary  dipolar BEC soliton
of energy  $E\equiv E_{\mathrm{kin}}+ E_{\mathrm{trap}}+ E_{\mathrm{int}}$
and chemical potential $\mu\equiv E_{\mathrm{kin}}+ E_{\mathrm{trap}}+ 2E_{\mathrm{int}}$
are obtained by solving 
Eqs. (\ref{f3}) and (\ref{f4}) for $\ddot w_\rho=\ddot w_z=0$.

\section{Numerical Results}
We perform    numerical simulation of the 3D GP equation (\ref{gp3d})
using   the split-step  Crank-Nicolson 
method  \cite{Muruganandam2009}.  
In the presence of the dipolar term the GP equation is integro-differential involving 
partial derivatives. 
The dipolar term is
treated by  fast Fourier transformation  \cite{jb}.
The   error of the reported numerical results 
is less than 1 $\%$.   

\begin{figure}
\begin{center}
\includegraphics[width=0.9\linewidth,clip]{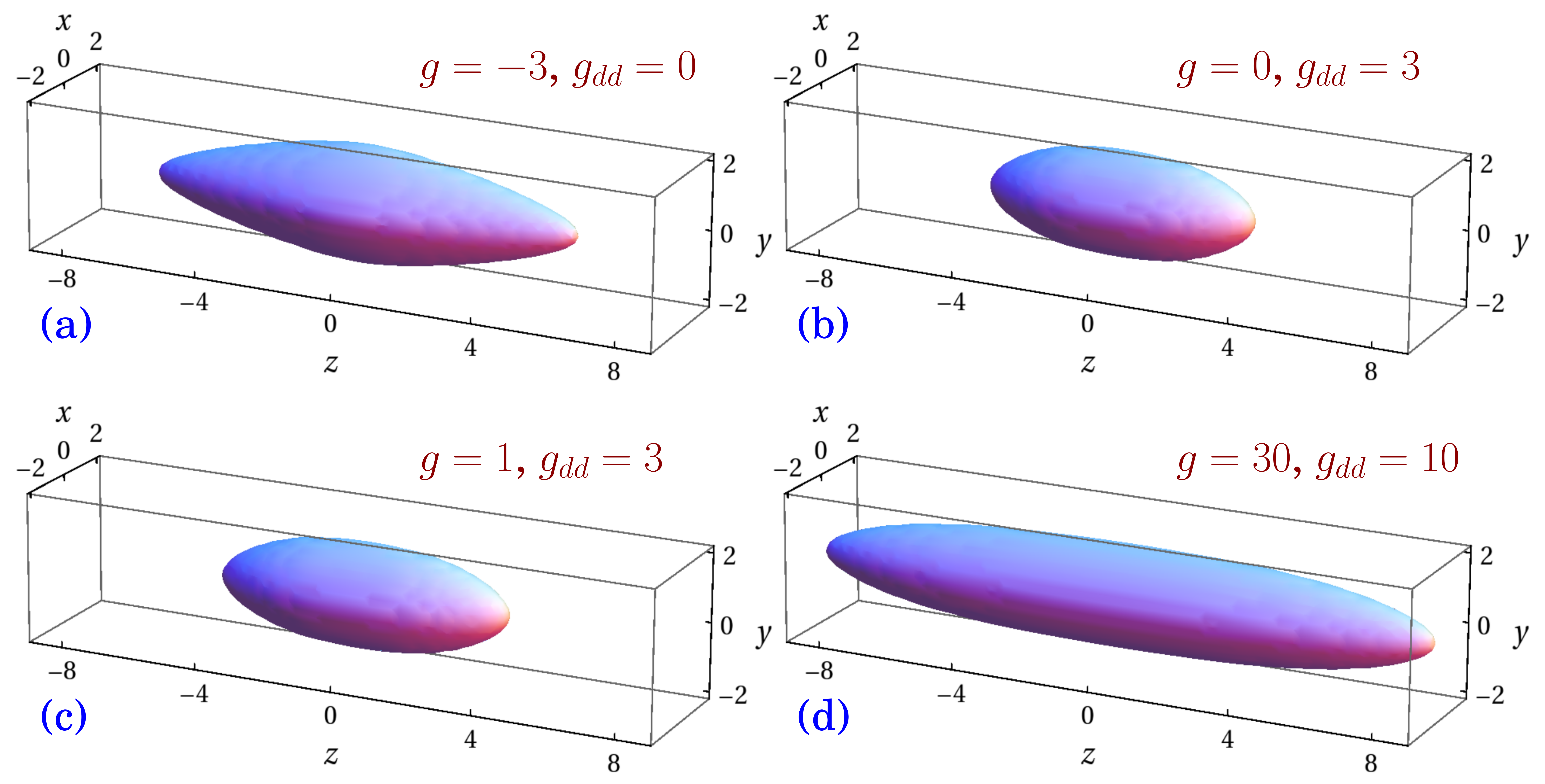}
\end{center}

\caption{(Color online) (a) Numerical 3D contour of a  nondipolar BEC
soliton of 1000 atoms with the scattering length tuned to
$a=-4.5a_0$ $ (g\equiv 4\pi a N=-3,gdd\equiv 3a_{dd}N=0).$
 The same for a dipolar BEC soliton of 1260 $^{52}$Cr atoms with scattering
length $a$ tuned to (b) $a=0$ ($g=0,g_{dd}=3$) and (c) $a=1.2a_0$ $(g=1,g_{dd}=3)$.
(d) The same for a dipolar BEC soliton of 4200 $^{52}$Cr atoms with scattering
length $a$ tuned to $a= 10.75a_0$ ($g=30,g_{dd}=10$).  
The square OL potential  is
$V^{2D}_{\text{OL}}=
-2\cos(2x)-2\cos(2y)$.
 The density
$|\phi({\bf r})|^2$ on the
contour is  0.001.
}
\label{fig1}
\end{figure}

First, we present the results for a single 1D dipolar BEC soliton on the square lattice.
In this case we take, for the weak 2D square OL, $V_0=2$.
 We present  the 3D contour   of    density  $|\phi({\bf r})|^2$ of the 
BEC  soliton  in Fig. \ref{fig1} for (a)      $g=-3$, $g_{dd}=0$,
(b)
  $g=0$, $g_{dd}=3$, (c)  $g=1$, $g_{dd}=3$, 
and (d)    $g=30$, $g_{dd}=10$
with  the  2D square  OL
$V^{2D}_{\text{OL}}=-2\cos(2x)-2\cos(2y)$, where $g \equiv 4\pi N a$ is 
the contact-interaction 
nonlinearity  and $g_{dd}\equiv
3Na_{dd}$ the dipolar nonlinearity.
The corresponding values of number of $^{52}$Cr atoms ($a_{dd}=15a_0$ with $a_0$ the Bohr 
radius \cite{lewen,pfau}) 
and the scattering lengths are given in the 
figure caption. 
 The density profiles are distinct in the four cases 
according to the net attraction in the system. The BEC soliton of Fig. \ref{fig1} (b) with no 
atomic repulsion ($g=0$) is the most attractive of the four corresponding to a small size, 
whereas the  BEC soliton of Fig. \ref{fig1} (d) with largest 
atomic repulsion ($g=30$) is the least attractive of the four corresponding to a large size.
In Fig \ref{fig1} (a), the BEC soliton is stabilized solely by atomic attraction ($g_{dd}=0$), and 
in Fig.  \ref{fig1} (b), the stabilization is achieved solely by dipolar interaction ($g=0$). 
 The 
numerical 
energy 
and root-mean-square (rms) sizes   of the  
BEC solitons of
 Fig. \ref{fig1} are shown in Table 
1 together 
with  the variational results.

\begin{table}
\label{I}
\caption{Numerical ($n$) and variational ($v$) energy and 
rms sizes, normal-mode frequencies   $E,\langle x\rangle, \langle y\rangle, \langle z\rangle$, $\Omega_\rho$
and $\Omega_z$, respectively, 
of  dipolar BEC solitons of Fig. \ref{fig1} on square OL. }
\centering
\label{table:1}
\begin{tabular}{lrrcccccc}
\hline
&$g$ & $g_{dd}$    & $E$  &  $\langle x\rangle$  & $\langle y\rangle$  &
  $\langle z\rangle$ & $\Omega_z$& $\Omega_\rho$ \\
\hline
$n$& $-3$ &  $0$ & $-1.512$ & $0.57 $ & $0.57 $ & $1.61 $ & $0.115$ & $3.453$ \\
$v$& $-3$ &  $0$ & $-1.489$ & $0.456$ & $0.456$ & $1.544$ & $0.203$ & $4.090$ \\
$n$& $ 0$ &  $3$ & $-1.802$ & $0.388$ & $0.388$ & $1.03 $ &  --     &   --    \\
$v$& $ 0$ &  $3$ & $-1.729$ & $0.395$ & $0.395$ & $1.140$ & $0.703$ & $4.332$ \\
$n$& $ 1$ &  $3$ & $-1.731$ & $0.413$ & $0.413$ & $1.21 $ & $0.614$ & $3.914$ \\
$v$& $ 1$ &  $3$ & $-1.694$ & $0.410$ & $0.410$ & $1.292$ & $0.566$ & $4.278$ \\
$n$& $30$ & $10$ & $-1.576$ & $0.488$ & $0.488$ & $3.03 $ & $0.154$ & $3.607$ \\
$v$& $30$ & $10$ & $-1.554$ & $0.445$ & $0.445$ & $3.164$ & $0.164$ & $4.130$ \\
\hline
\end{tabular}
\end{table}

One can have a 1D {\it nondipolar}  ($g_{dd}=0$) 
BEC soliton 
on the 2D {\it square} OL for $0>g  >-g_{\text{crit}}$, where the 
numerical estimate 8.60
of    $g_{\text{crit}}$ should be contrasted 
with the variational result of  6.16. As $g=4\pi N a$, these correspond 
to the critical values $|Na|_{\text{variational}}= 0.490 $ and 
$|Na|_{\text{numerical}}= 0.684 $.  The present numerical critical value is 
surprisingly close to the   
following critical value  when the 2D OL trap in the $x$-$y$ plane is replaced 
by the harmonic trap $V=\rho^2/2$:  
$|Na|_{\text{numerical}}= 0.676 $ \cite{crit}. This shows the similar nature of the 
two nondipolar BEC solitons.

\begin{figure}[!t]
\begin{center}
\includegraphics[width=0.9\linewidth,clip]{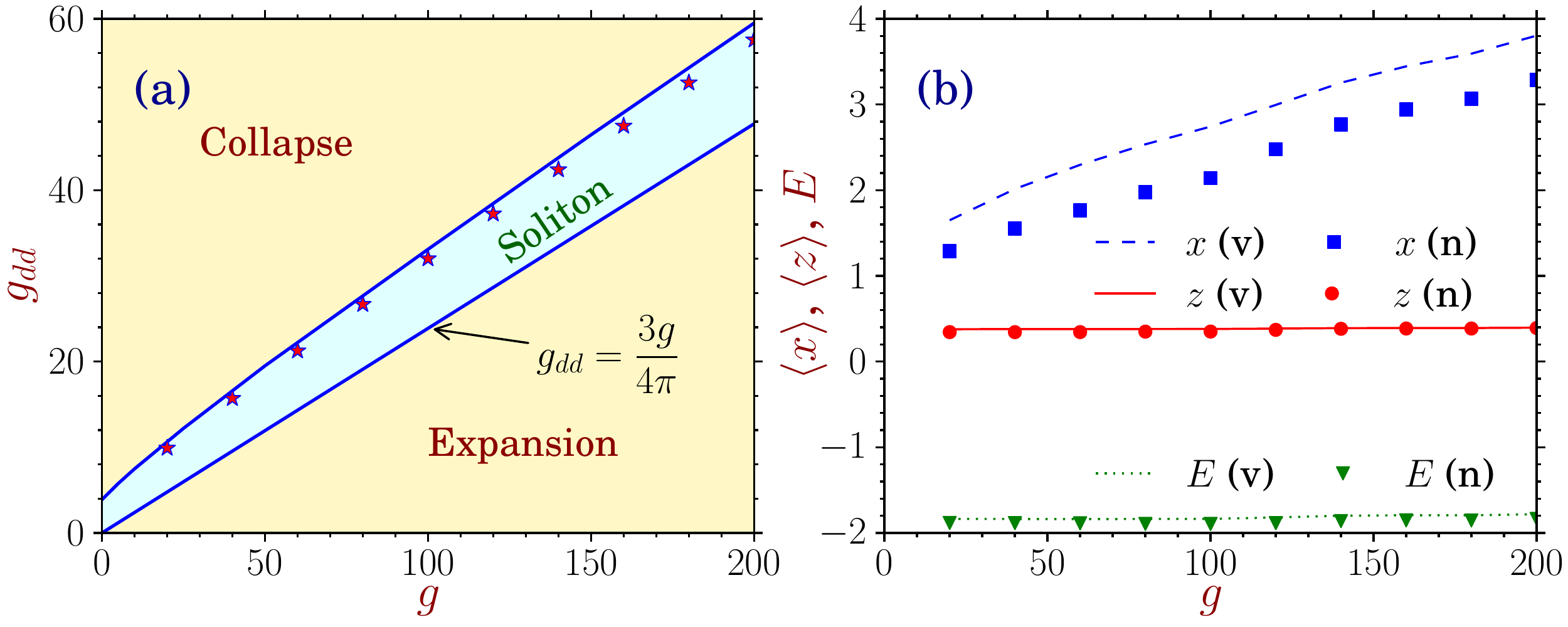}
\end{center}

\caption{(Color online)  (a) The phase plot of $g_{dd}$ versus $g$ from variational analysis showing 
the region of stable dipolar BEC  soliton formation on square OL with $V_0 =2$. The $\star$'s denote the numerical points showing 
the stable-collapse boundary. (b) The numerical ($n$) and variational ($v$) rms sizes and energy versus $g$ for 
$g_{dd}$ corresponding to the  $\star$'s in (a). 
}
\label{fig2}
\end{figure}

For a fixed nondipolar nonlinearity $g$,
an 1D dipolar BEC   soliton  ($g_{dd}>0$) on the 2D {\it square} OL
can be stabilized for 
$g_{dd}^{\text{crit1}}>g_{dd}>g_{dd}^{\text{crit2}}$, where for 
$g_{dd}<g_{dd}^{\text{crit2}}$ there is not sufficient net attraction 
 and the BEC soliton expands to infinity and for $g_{dd}>g_{dd}^{\text{crit1}}$
there is too much net attraction leading to collapse. The domain of stable soliton 
in this case is shown in the $g_{dd}$ versus $g$ phase plot in 
Fig. \ref{fig2} (a), where the two lines are the variational 
boundaries between stable BEC soliton    and collapse and that between stable BEC soliton   and expansion.  
Of these two lines, the lower boundary between stable BEC soliton and expansion can be analytically 
obtained from the variational equations (\ref{f3}) and (\ref{f4}) for $\ddot \omega_z=\ddot \omega_\rho =0$.
In this limit the BECs will be infinitely large accommodating an infinite number $N$ of atoms
and in order that Eqs.    (\ref{f3}) and (\ref{f4}) yield finite numbers for $N\to \infty$ 
one must have $[a-a_{dd}h(\kappa)]=[a-a_{dd}g(\kappa)/2]=0$, so that $h(\kappa)=g(\kappa)/2$ 
with the solution $\kappa=0$, while $h(\kappa=0)=1$. Consequently, this boundary is defined by 
$a=a_{dd}$ or $g_{dd}=3g/(4\pi)$ corresponding to the 
lower line in Fig. \ref{fig2} (a).   In this figure the $\star$'s denote the numerically calculated boundary 
between collapse and stability. In Fig. \ref{fig2} (b) we plot the numerical and variational sizes and energies 
versus $g$ for $g_{dd}$ corresponding to the $\star$'s in Fig. \ref{fig2} (a).

To perform a linear stability analysis \cite{stability} of the BEC solitons on the square OL
and obtain the normal-mode frequencies, we note 
that Eqs.   (\ref{f3}) and (\ref{f4}) can be rewritten as \cite{luca}
\begin{eqnarray}
\ddot w_\rho = - \frac{\partial U }{\partial w_\rho}, \quad \ddot w_z = - \frac{\partial U }{\partial w_z},
\end{eqnarray}
where $U$ is the linearized effective potential. The squares of the normal-mode frequencies $\Omega_z$ and  $\Omega_\rho$
along axial $z$ and transverse directions, respectively, 
   are the eigenvalues of the 
 eigenfunction-eigenvalue problem for the Hessian matrix $\Lambda_{ij}=\partial^2 U/(\partial w_i\partial w_j)|_{w_z=w_{z}^\star,
w_\rho=w_{\rho }^\star}$ where $i,j=z,\rho$,
where $w_{z}^\star$ and $w_{\rho }^\star$ are the stationary solutions of  Eqs.   (\ref{f3}) and (\ref{f4}) obtained by
setting $\ddot w_\rho=\ddot w_z=0$ \cite{luca}. If these frequencies are real, 
stable oscillation of the widths is assured, whereas imaginary or complex frequencies imply exponential increase or decrease 
of the widths upon small perturbation corresponding to unstable states.   These frequencies for the states exhibited in Fig. \ref{fig1} are shown in Table 1. 
The real     
  normal-mode frequencies of the BEC solitons of Fig. \ref{fig1}, displayed in Table 1, 
guarantee their stability.
The   numerical frequencies in Table 1
 were calculated from the small oscillations of the widths 
in real-time propagation for a long    time.  
The variational analysis provides a qualitative understanding of many features of the BEC  
soliton 
including its stability and the normal-mode frequencies.

\begin{figure}[!t]
\begin{center}
\includegraphics[width=.45\linewidth,clip]{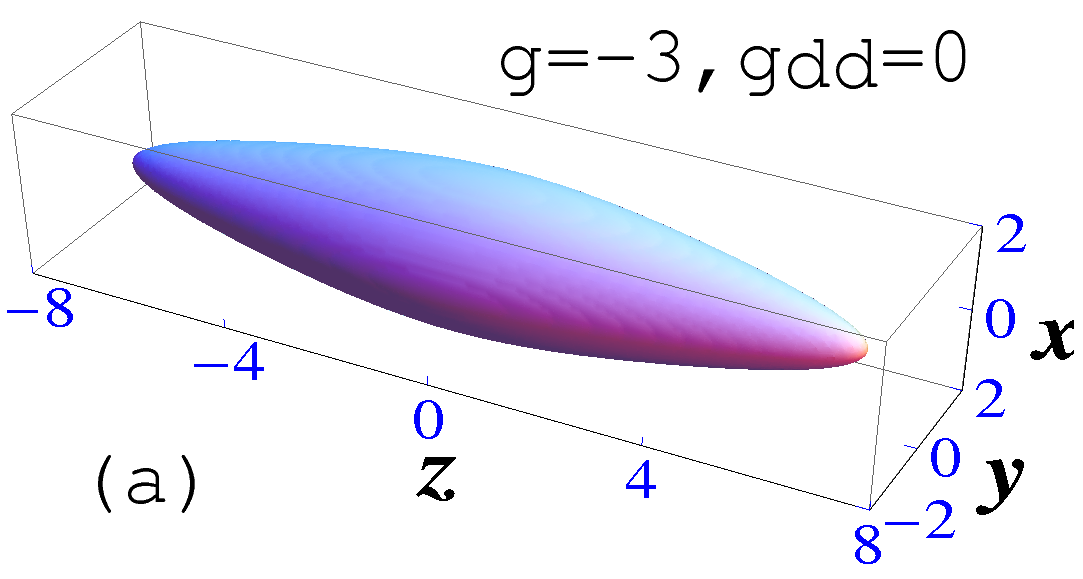}
\includegraphics[width=.45\linewidth,clip]{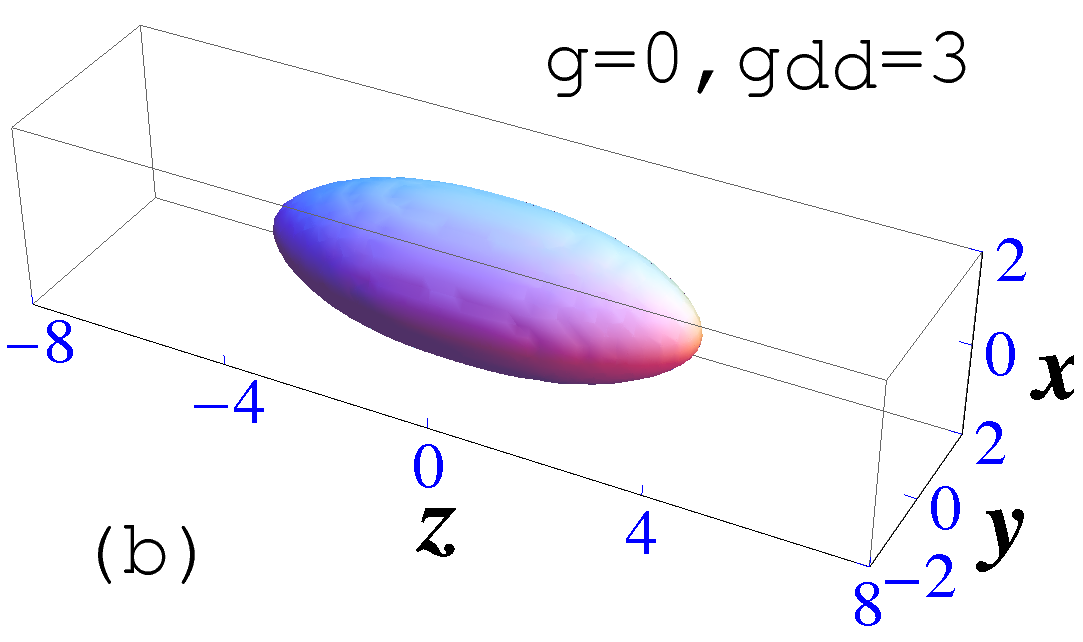}
\includegraphics[width=.45\linewidth,clip]{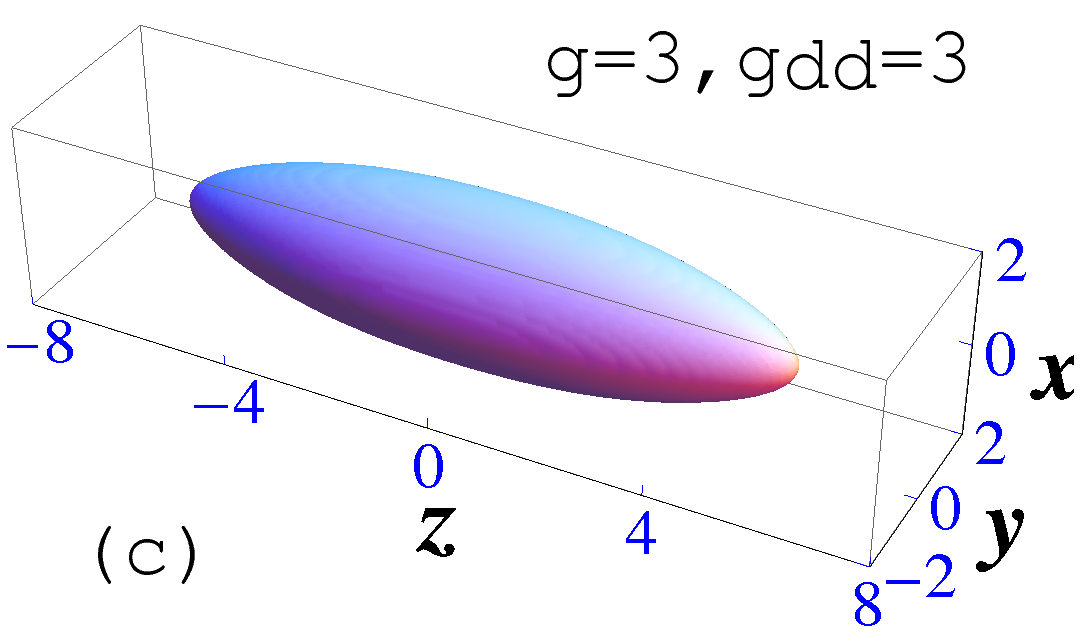}
\includegraphics[width=.45\linewidth,clip]{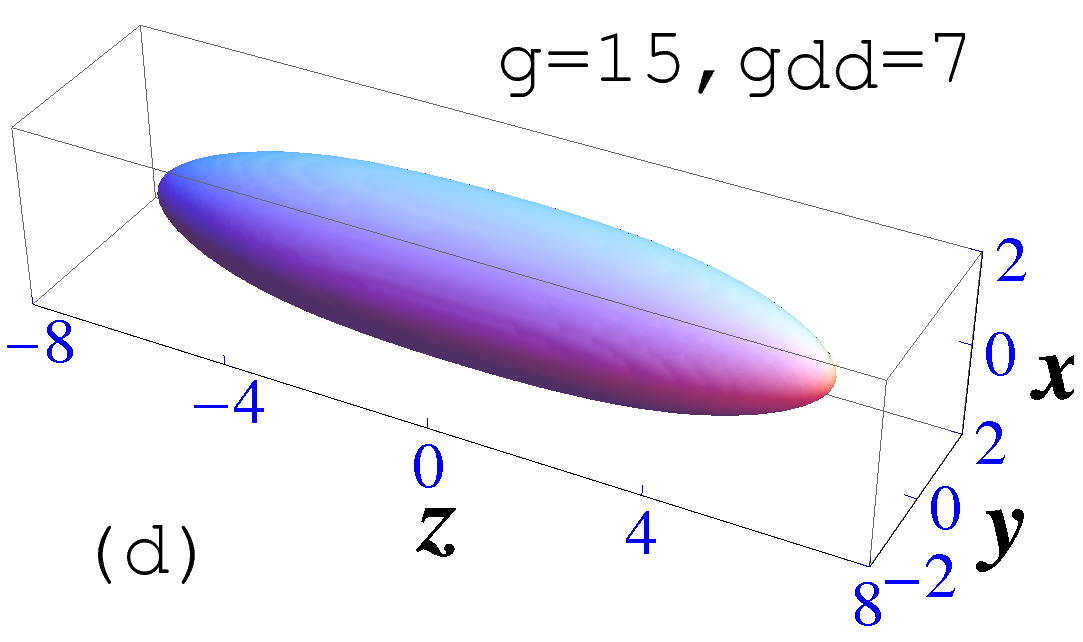}
\end{center}

\caption{(Color online)  Numerical  3D contour of the dipolar 
BEC solitons  for (a) $g=-3$ and $g_{dd}=0$,  (b) $g=0$ and $g_{dd}=3$,  (c) $g=3$ and $g_{dd}=3$,
 and (d) $g=15$ and $g_{dd}=7$ on the triangular OL 
$V^{2D}_{\text{OL}}= -\cos(2x)-\cos(x+\sqrt 3 y)-\cos(x-\sqrt 3 y)$.  
 The density 
$|\phi({\bf r})|^2$ on the 
contour is  0.001.
}
\label{fig3}
\end{figure}

Next, we present the results for a  1D dipolar BEC soliton on the {\it triangular} OL
in brief.
 We take, for the weak 2D triangular  OL, $V_0=1$.
 We present  the 3D contour   of    density  $|\phi({\bf r})|^2$ of the stable
BEC  soliton  in Fig. \ref{fig3} for (a)      $g=-3$, $g_{dd}=0$,
(b)
  $g=0$, $g_{dd}=3$, (c)  $g=3$, $g_{dd}=3$, 
and (d)    $g=15$, $g_{dd}=7$
with  the  2D triangular  OL
$V^{2D}_{\text{OL}}=-\cos(2x)-\cos(x+\sqrt 3y)-\cos(x-\sqrt 3 y)$. 
The numerical energy and  
rms sizes of the  solitons of Fig. \ref{fig3} are shown in Table
2 together with the variational results.   The  normal-mode frequencies calculated from the 
linear stability analysis \cite{stability}, as shown in Table 2, indicate the stability of the solitons.

\begin{table}
\label{II}
\caption{Numerical ($n$) and variational ($v$) energy and 
rms sizes, normal-mode frequencies   $E,\langle x\rangle, \langle y\rangle, \langle z\rangle$, $\Omega_\rho$
and $\Omega_z$, respectively,  
of  dipolar BEC solitons of Fig. \ref{fig3} on triangular OL. }
 
\centering
\label{table:2}
\begin{tabular}{lrrcccccc}
\hline
&$g$ & $g_{dd}$    & $E$  &  $\langle x\rangle$  & $\langle y\rangle$  &
  $\langle z\rangle$ & $\Omega_z$& $\Omega_\rho$ \\
\hline
 $n$ & $-3$ & $0$ & $-0.876$ & $0.678$ & $0.678$ & $2.07 $ & $-$     & $-$     \\ 
 $v$ & $-3$ & $0$ & $-0.856$ & $0.503$ & $0.503$ & $1.881$ & $0.138$ & $3.286$ \\
 $n$ &  $0$ & $3$ & $-1.082$ & $0.443$ & $0.443$ & $1.22 $ & $-$     & $-$     \\
 $v$ &  $0$ & $3$ & $-1.046$ & $0.443$ & $0.443$ & $1.315$ & $0.519$ & $3.533$ \\
 $n$ &  $3$ & $3$ & $-0.965$ & $0.500$ & $0.500$ & $1.8  $ & $-$     & $-$     \\
 $v$ &  $3$ & $3$ & $-0.944$ & $0.478$ & $0.478$ & $1.855$ & $0.275$ & $3.395$ \\
 $n$ & $15$ & $7$ & $-1.002$ & $0.483$ & $0.483$ & $2.247$ & $-$     & $-$     \\
 $v$ & $15$ & $7$ & $-0.978$ & $0.469$ & $0.469$ & $2.1  $ & $0.262$ & $3.429$ \\
\hline
\end{tabular}
\end{table}

\begin{figure}[!b]
\begin{center}
\includegraphics[width=0.9\linewidth,clip]{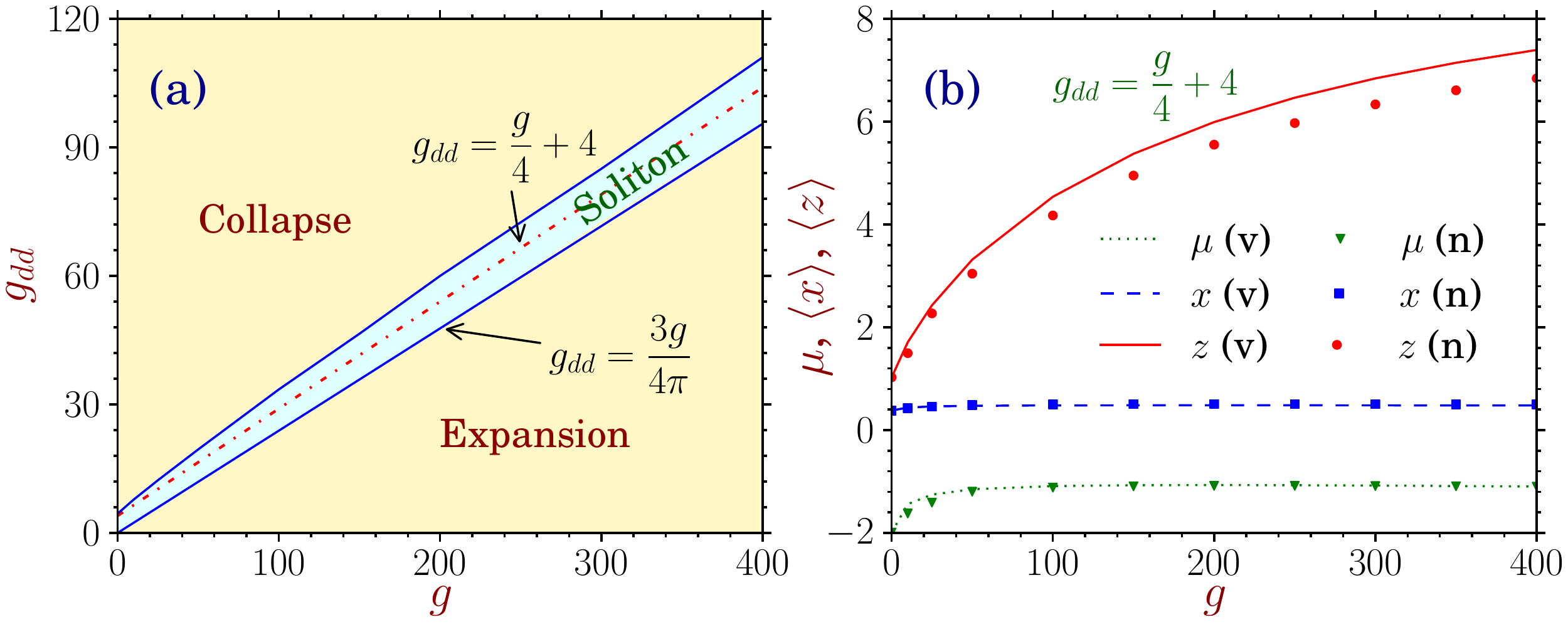}
\end{center}

\caption{(Color online)  (a) The phase plot of $g_{dd}$ versus $g$ from variational analysis showing 
the region of stable dipolar BEC soliton   on triangular OL  for $V_0 =1$. 
 (b) The numerical (n) and variational (v) rms sizes and chemical potential  versus $g$ corresponding 
to the line 
$g_{dd}=4+g/4$ in (a). 
}
\label{fig4}
\end{figure}

The domain of stable dipolar BEC solitons on triangular OL is illustrated in Fig. \ref{fig4} (a)
in a phase plot of the nonlinearities $g$ and $g_{dd}$ as obtained from the variational equations
(\ref{f3}) and (\ref{f4}). Again there is a domain of stable soliton 
 between a domain of collapse and of expansion. The collapse takes place for  too large a 
value of dipolar nonlinearity $g_{dd}$ and expansion for too small a value of dipolar nonlinearity. 
A moderate value of the dipolar nonlinearity leads to stable solitons. The boundary between stable
soliton  and expansion is again given by the analytic formula $g_{dd}= 3g/(4\pi)$. In Fig. 
\ref{fig4} (b) we compare the variational and numerical chemical potential and rms sizes $\langle x
\rangle$ and $\langle z \rangle$ along the line $g_{dd}= 4+g/4$ covering the whole domain of soliton 
formation $400> g>0$ shown in Fig. \ref{fig4} (a).

\begin{figure}[!ht]
\begin{center}
\includegraphics[width=0.9\linewidth,clip]{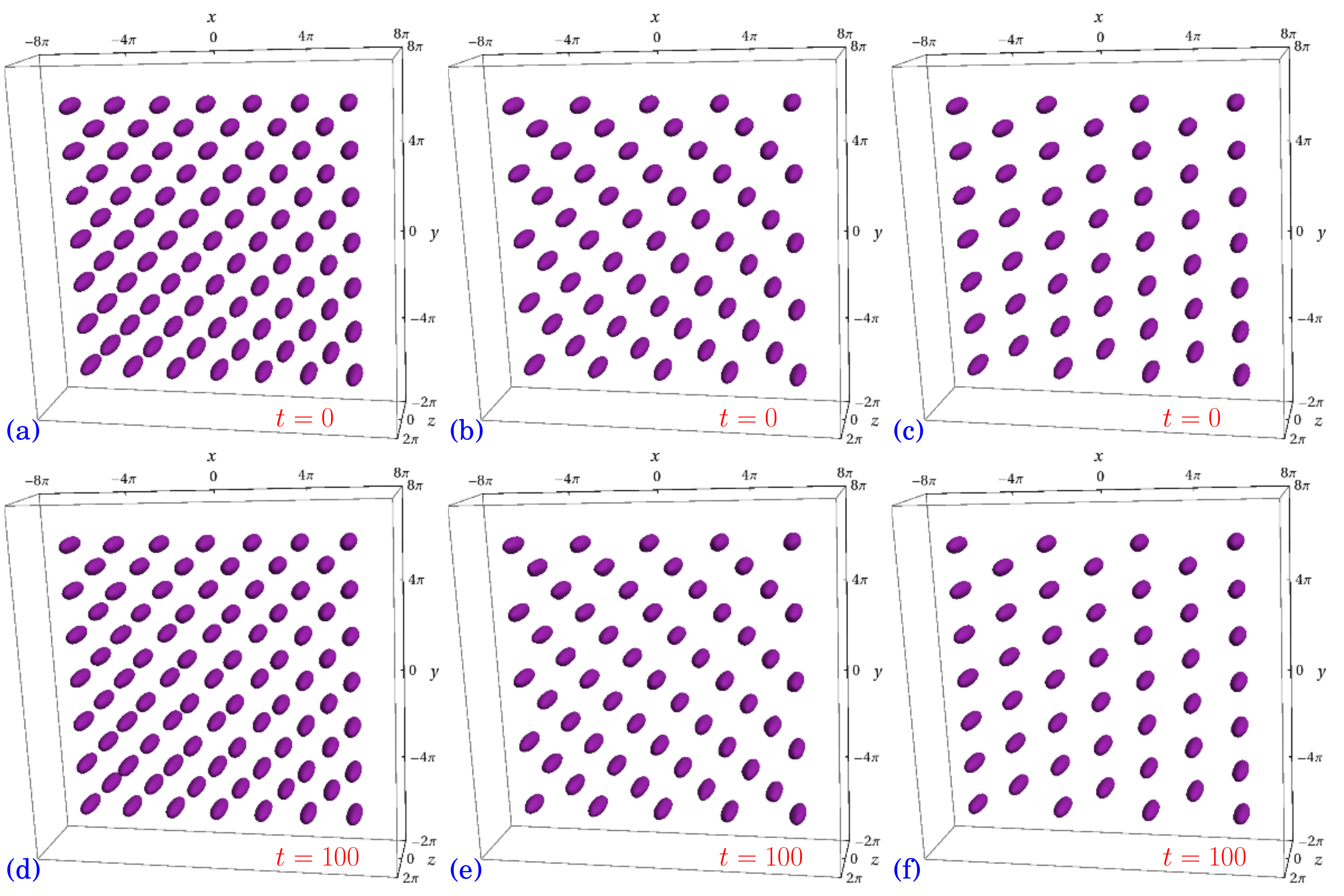}
\end{center}

\caption{(Color online) Stable array of tiny dipolar BEC solitons, each with
$g=1, g_{dd}=3$ on the  square OL with $V_0=2$  in the 
(a)   checkerboard, (b) stripe, and (c) star configurations with filling factors of 1/2,
1/3, and 1/4, respectively at $t=0$. The same arrays after real-time propagation at 
$t=100$ are shown in (d), (e), and (f), respectively.
}
\label{fig5}
\end{figure}

One interesting aspect of studying 1D BEC soliton   on 2D lattice is to consider   an 
array of many tiny 1D BEC  soliton droplets distributed in 2D OL sites so that a periodic 
distribution of matter simulating a 2D solid in condensed-matter physics with long-range 
inter-site interaction is obtained. To achieve this, we consider  tiny dipolar BEC
solitons on square OL with $g=1$ and 
$g_{dd}=3$ and distribute these on different   sites and study the stability of such an 
array by solving the GP equation  by real-time 
propagation. We find that such an array is always unstable due to long-range 
inter-site interaction, if two solitons are placed 
on neighboring sites along one of the axes $-$ $(x,y)$. However, the array is stable 
if they are placed along the diagonal directions. With this information, we see that a
stable periodic 2D pattern of tiny solitons is obtained if the occupation of neighboring 
sites is avoided. At the maximum of a filling factor of 1/2, the stable checkerboard 
configuration is displayed in Fig. \ref{fig5} (a), where the soliton droplets are put diagonally 
on the black or white spots of a chess board. After real-time evolution of the GP equation, 
the final array at $t=100$ is displayed in  Fig. \ref{fig5} (d). It is interesting to 
note that this checkerboard pattern is a  stable  Mott insulator state obtained by solving 
the Bose-Hubbard model on a 2D lattice with repulsive long-range dipolar interaction 
\cite{barbara}. The other Mott states obtained there are the  stripe and star configurations 
with 1/3 and 1/4 filling of sites, respectively. We considered such configurations with the 
present tiny dipolar BEC  soliton droplets. By reducing the filling factor from 1/2 to 1/3 or 1/4 
we have a lower occupation of sites and hence a lower inter-site interaction. As such arrays
get destroyed due to the long-range inter-site interaction,  a reduced inter-site interaction 
implies that the stripe and star configurations of the BEC soliton droplets are stable. We 
established the stability of the stripe and star configurations using real-time propagation of 
the GP equation. In Figs. \ref{fig5} (b) and (c)  we show the initial  stripe and star configurations
of the tiny BEC droplets at $t=0 $,  and in Figs. \ref{fig5} (e) and (f)  the same 
after real-time propagation at $t=100$. The initial and final configurations are practically 
indistinguishable, demonstrating the stability.

\section{Summary and conclusion}
To summarize, using the mean-field 3D GP equation 
we demonstrated  a stable dipolar BEC  soliton polarized 
along the axial $z$ direction 
on a weak square or triangular 
2D OL in the orthogonal $x$-$y$ plane. We considered identical OL strengths along different 
directions and considered a  Lagrangian variational analysis of the GP equation with a Gaussian 
ansatz in addition to the numerical solution of the same using the split-step Crank-Nicolson method.
The stabilization of the BEC was established by the linear stability analysis \cite{stability}. 
The widths and energies obtained from the numerical solution of the GP equation are in agreement 
with the corresponding variational results.

We also considered stable 2D arrays formed by arranging identical tiny dipolar BEC solitons on different 
sites of a 2D square OL.  Such an array is unstable if any two adjacent OL sites are occupied. The simplest stable periodic 
2D array of superfluid dipolar BEC solitons, known as the checkerboard configuration,  
emerges 
if the sites are arranged diagonally with a filling factor of 1/2.  
Similar  2D arrays, known as  stripe and star configurations for filling factors 1/3 and 1/4, respectively 
are also found to be stable. Previously, in the study of ultra-cold dipolar atoms on strict 
2D lattice using the Monte Carlo simulation  of the 2D Bose-Hubbard model,
such stable arrays of Mott insulator states emerged at filling factors of 1/2, 1/3, and 1/4  \cite{barbara}. Similar stable 
configurations obtained in the mean-field GP and field-theoretic Bose-Hubbard approaches possibly indicate 
the general stability property  of such structures under the long-range repulsive 
dipolar interaction in the $x$-$y$ plane.

\section*{Acknowledgements}

We thank
FAPESP (Brazil),   CNPq (Brazil),    DST (India),   and CSIR  (India)
for partial support.


\section*{References}

\begin{thebibliography}{99}
 
  
 

\bibitem{rmp} F. Dalfovo, S. Giorgini, L. Pitaevskii,  S. Stringari,  Rev. Mod. Phys. {  71} (1999) 463.

\bibitem{sw} F. S. Cataliotti  {\it et al.}, Science {  293} (2001)  843.


\bibitem{band}J. Heinze, S. G\"otze, J. S. Krauser, B. Hundt, N. Fl\"aschner, D.-S. L\"uhmann, C. Becker, and K. Sengstock, 
\prl {  107} (2011) 135303.


\bibitem{lewen}M. Lewenstein, A. Sanpera, V. Ahufinger, B. Damski, A. Sen De, and U. Sen,    Adv. Phys. {  56} (2007) 243.



 

\bibitem{pfau}T. Koch {\it et al.},
{ Nature Phys.} {  4} (2008) 218;\\
T.
  Lahaye  {\it et al.}, 
  { Nature} {  448}  (2007) 672;\\
  T. Lahaye {\it et al.} 2008 
\prl
  {  101} (2008) 080401;\\
A.
Griesmaier {\it et al.},
\prl {  97} (2006) 
250402.




\bibitem{rpp}T.  Lahaye {\it et al.},
 { Rep. Prog. Phys.}
{  72} (2009) 126401.

 

 

\bibitem{dy1}  
M. Lu, S. H. Youn,  B. L. Lev, 
Phys. Rev. Lett. {  104} (2010) 063001;\\ 
J. J.
McClelland,  
J. L.  Hanssen,  \prl
{  96} (2006) 143005;\\
S. H. Youn, M. W. Lu, U. Ray,   B. V. Lev, 
 \pra  {  82} (2010) 043425;\\
M. Lu, N. Q. Burdick, Seo Ho Youn,   B. L. Lev,
\prl {  107}  (2011) 190401.



\bibitem{stab}N. G.  Parker, C. Ticknor, A. M. Martin,    D. H. J. O'Dell,
  \pra {  79} (2009) 013617;\\
M. Asad-uz-Zaman,   D. Blume, 
 Phys. Rev. A {  80} (2009) 053622;\\
R. M. Wilson, S. Ronen,   J. L. Bohn, \pra {  80} (2009) 023614.



\bibitem{stability}See, for example, 
J. E. Howard,      R. S. MacKay,  
Phys. Lett. {  A122} (1987) 331;\\
M. Tabor, {\it Chaos and Integrability in Nonlinear Dynamics: An Introduction},  (New York: Wiley, pp. 20-31, 1989).



\bibitem{luca}E. Cerboneschi, R. Mannella, E. Arimondo,   L. Salasnich, Phys. Lett. A {  249} (1998)  495;\\
L. Salasnich, Int. J. Mod. Phys. B {  14} (2000) 1;\\
S. Stringari, \prl {  77} (1996) 2360. 
 

\bibitem{barbara} B. Capogrosso-Sansone, C. Trefzger, M. Lewenstein,  P. Zoller,   G. Pupillo, \prl 
{  104} (2010) 125301.





\bibitem{Tikhonenkov2008}
I. Tikhonenkov, B. A. 
  Malomed,   A. Vardi, \prl  
 {  100} (2008) 090406.
 


\bibitem{Pedri2005} 
R. Nath, P. 
 Pedri,     L. Santos,   
{\prl}  {  102} (2009) 050401;\\
P. Pedri,   L. Santos, Phys. Rev. Lett.
{  95} (2005) 200404.



\bibitem{adhisol}L. E. Young-S, P. Muruganandam,   S. K. Adhikari, J. Phys. B {  44} (2011) 101001.

 

\bibitem{salerno}B. B. Baizakov, B. A. Malomed, and M. Salerno, Phys. Rev. A
{\bf 70,} 053613 (2004).


\bibitem{jb}K. 
 G\'oral,    L.  Santos,   \pra  {  66} (2002) 023613;\\
S. Yi,    L. You,   \pra 
{  63} (2001) 053607;\\
S. Yi,    L. You, 
 \prl
 {  92} (2004) 193201.




\bibitem{Muruganandam2009}
P. Muruganandam,    S. K.  Adhikari, { Comput. Phys.
  Commun.} {  180} (2009) 1888;\\ P. Muruganandam,    S. K.  Adhikari,
J. Phys. B {  36} (2003) 2501;\\
S. K.  Adhikari,   P. Muruganandam,
J. Phys. B {  35} (2002) 2831.

  
 
 \bibitem{crit} S. K. Adhikari, New J. Phys. { 5} (2003) 137;\\
V.  M. P\'erez-Garc\'ia, H. Michinel,   H. Herrero,  Phys. Rev. A { 57} (1998) 3837;\\
L. Salasnich, A. Parola,    L. Reatto,  Phys. Rev. A { 66} (2002) 043603;\\
A. Gammal, L. Tomio,   T. Frederico,   Phys. Rev. A { 66} (2002) 043619.
 

 \end{thebibliography}
\end{document}